# Edge-Grafted Molecular Junctions between Graphene Nanoplatelets: Applied Chemistry to Enhance Heat Transfer in Nanomaterials

*M.Mar Bernal, Alessandro Di Pierro, Chiara Novara, Fabrizio Giorgis, Bohayra Mortazavi, Guido Saracco and Alberto Fina\**

Dr. M.M. Bernal, A. Di Pierro, Prof. A. Fina
Dipartimento di Scienza Applicata e Tecnologia
Politecnico di Torino, Alessandria, 15121, Italy
E-mail: alberto.fina@polito.it

Dr. C. Novara, Prof. F. Giorgis, Prof. G. Saracco
Dipartimento di Scienza Applicata e Tecnologia
Politecnico di Torino, Torino, 10129, Italy

Dr. B. Mortazavi
Institute of Structural Mechanics, Bauhaus-Universität Weimar
Marienstr. 15, D-99423 Weimar, Germany

The edge-functionalization of graphene nanoplatelets (GnP) was carried out exploiting diazonium chemistry, aiming at the synthesis of edge decorated nanoparticles to be used as building blocks in the preparation of engineered nanostructured materials for enhanced heat transfer. Indeed, both phenol functionalized and dianiline-bridged GnP (GnP-OH and E-GnP, respectively) were assembled in nanopapers exploiting the formation of non-covalent and covalent molecular junctions, respectively. Molecular dynamics allowed to estimate the thermal conductance for the two different types of molecular junction, suggesting a factor 6 between conductance of covalent *vs.* non-covalent junctions. Furthermore, the chemical functionalization was observed to drive the self-organization of the nanoflakes into the nanopapers, leading to a 20% enhancement of the thermal conductivity for GnP-OH and E-GnP while the cross plane thermal conductivity was boosted by 150% in the case of E-GnP. The application of chemical functionalization to the engineering of contact resistance in nanoparticles network was therefore validated as a fascinating route for the enhancement of heat exchange efficiency on nanoparticle networks, with great potential impact in low-temperature heat exchange and recovery applications.

## 1. Introduction

The fast-growing development of modern technologies made efficient heat dissipation extremely important to the performance, lifetime and reliability of electronic and optoelectronic devices. In this regard, there is an urgent need for the development of flexible and lightweight thermally conductive materials to improve the thermal management efficiency of such systems. Graphene sheets, owing to their exceptional mechanical and electrical properties and high intrinsic thermal conductivity,[1] are of great interest to be used as building blocks to create macroscopic assembled materials with unique properties.[2] While graphene, defined as a single layer $sp^2$ carbon, currently remains of insufficient availability for the use in large scale bulk applications, graphene related materials (GRM), including reduced graphene oxide, few layer graphene, multilayer graphene and graphene nanoplatelets, represent the state of the art materials for the exploitation into thermally conductive applications. Specifically, large-area freestanding "paper-like" materials made from GRM have emerged as promising materials to address heat dissipation problems in practical applications. [2d, 3]

GRM nanopapers show strongly anisotropic thermal conductivity between the in-plane and cross-plane directions, [3c] reflecting the strong covalent $sp^2$ bonding between carbon atoms on graphene sheets, and the weak van der Waals interactions between them. [4] Engineering the heat conduction in GRM nanopapers is non-trivial. On the one hand, the thermal conductivity depends on the quality of the individual GRM sheets, *i.e.* the number of defects per unit area and the aspect ratio which determine the phonon transmission on the individual nanoflake. [2d, 5] Higher concentration of defects such as vacancies, inclusions, stacking defects, oxidized carbons or other functional groups are indeed well known to restrain the thermal conductivity of the individual nanoflakes. [6] On the other hand, the physico-chemical nature of the contact between the nanoflakes and the extension of the contact area inside the nanopaper determine the thermal contact resistance. [3a, 5b, 7] Previous experimental and theoretical studies have reported the dependency of the thermal conductivity with the average lateral size of the graphene, indicating that the heat conduction in graphene is limited by the phonon mean free path. [8] However, an increase in the average lateral size of few-layer-graphene flakes was also demonstrated to increase the thermal conductivity of the microlayer deposited on a polymer film, suggesting that the overall conductivity of the network is indeed limited by the contact resistance between the nanoflakes. [5b] It is worth noting that air cavities are typically obtained in the deposition of GRM nanoflakes, owing to the non-planarity of the nanoflakes and/or defects in stacking and orientation. The reduction of the number and total volume of the air cavities in the graphene assemblies obviously enhances the thermal contact between particles, thus improving the overall thermal conductivity. [7b] A simple way to reduce porosity of the deposition and enhance particle-particle contacts is by mechanical compaction via the application of uniaxial pressure on the deposition, which was demonstrated beneficial for the improvement of the thermal conduction. [5b] To maximise the thermal conductivity of graphene nanopapers, annealing at extremely high temperatures are typically carried out on the preformed papers, in order to restore complete $sp^2$ hybridization in the graphene sheets by removing oxidized groups and recombine structural defects [9] or by the formation of new $sp^2$ clusters. [10] Beside the increase in conductivity of the individual nanoflakes, [6a] defectiveness reduction upon thermal annealing may also affect the extent of $π$-$π$ interactions, finally improving the contact between adjacent nanoflakes. Furthermore, coalescence of overlapped sheets may occur at high annealing temperature, leading to the formation of extended polycrystalline layers. [11] Such approach endows graphene nanopapers with superior thermal conductivities in the in-plane direction but extremely low in the normal direction. Despite this may be fine when only considering heat spread over the nanopapers, the thermal transport along the cross-plane direction remains crucial to guarantee an efficient thermal contact between the GRM paper spreader and the heater and/or the heat sink.

To tailor the properties of nanopapers, the exploitation of molecular junctions between GRM nanoflakes is currently a promising possibility, yet experimentally challenging. In fact, while the conductance of molecular junctions wa widely studied by molecular dynamics [12] and Density Functional Theory, [13] the experimental exploitation of organic molecular junctions was only recently reported by Han *et al*. [14] for the thermal coupling of the graphene-graphene oxide and the graphene oxide-silica surfaces. In this work, both covalent and non-covalent molecular junctions were designed and synthetized to create GRM-based nanopapers with inherently low contact thermal resistance between nanoplatelets. Such molecular junctions were built at the edges of the nanoflakes by means of the controlled diazonium chemistry, allowing to preserve the defect-free $sp^2$ structure, as confirmed by XPS and Raman spectroscopy. Enhanced heat transfer performance of the nanopapers was assessed experimentally in both in-plane and cross-plane directions and interpretation of these results was further supported by molecular dynamics and finite element modelling.

## 2. Results and discussion

The objective of this study was to manufacture and validate molecular junctions between graphene nanoplatelets for the modulation of the thermal conductivities of GRM networks. In this framework, edge-selective functionalization is of utmost interest to obtain engineered nanoparticles able to further react or assemble in a finely controlled way, while preserving the high conductivity associated to the defect-free $sp^2$ structure of graphene. As a first step towards that target, we investigated the diazonium reaction conditions, to maximize the amount of chemical functions added at the edges of our graphene nanoplatelets (GnP) without affecting the conjugated $\pi$-system. The well-established procedure [15] for the in situ formation of the diazonium species from an anilinic compound, namely 4-aminophenol, in the presence of an alkyl nitrite was followed in this work (Scheme 1a and S1). Thermogravimetric analysis (TGA), X-ray photoelectron spectroscopy (XPS) and Raman spectroscopy were used to characterize the amount of species attached to GnP (Figure S1), their chemical composition (Figure S2, S3, S4, S5 and Table S1) and the microstructure changes of GnP due to the covalent functionalization (Figure S6). The optimized conditions for the edge-selective functionalization of our GnP using the diazonium chemistry were found to be 4 equiv. per C atom of the anilinic molecule for 24 h of reaction time (GnP-OH). Indeed, at lower concentration the degree of functionalization decreased, while at higher reaction times no improvements on the extent of the reaction were observed while the number of OH groups diminished.

The next step was to edge-link GnP to produce the molecular junction between nanosheets. To do so, 1,5-bis(4-aminophenyloxy)pentane (**3**) was used to establish an equivalent molecular junction in one-step (Scheme 1b), by the in situ formation of the diazonium salts of compound **3**, referred to as edge-linked GnP (E-GnP) (Scheme 1b). The reaction was carried out by adapting the optimized diazonium conditions established above for the edge-selective functionalization of our GnP with 4-aminophenol (See Experimental Section for details).

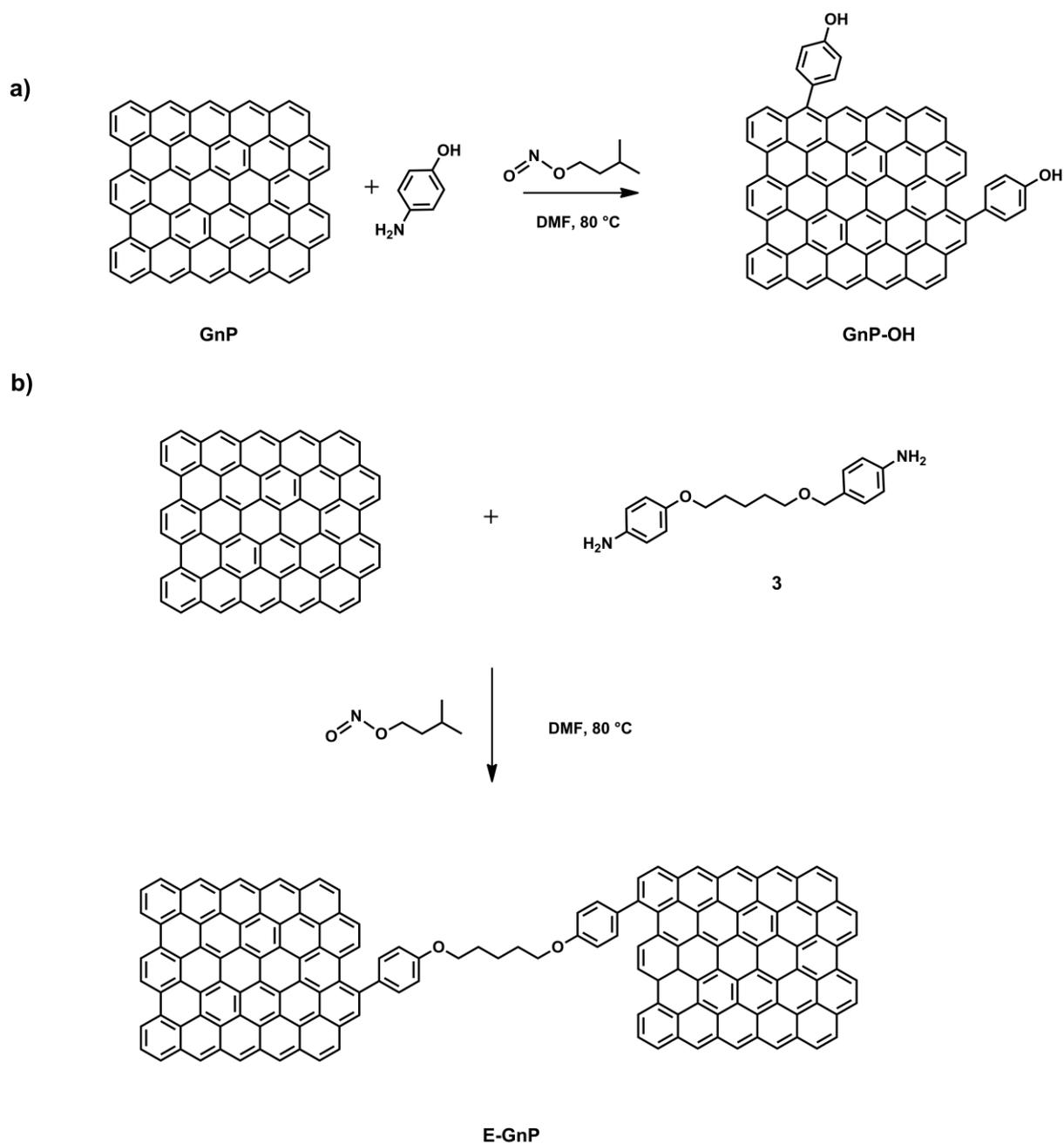

**Scheme 1.** a) Edge-selective functionalization with 4-aminophenol. b) Edge-linked GnP (E-GnP) with 1,5-bis(4-aminophenyloxy)pentane (**3**).

Evidences on the functionalization of GnP come from the XPS spectroscopic measurements. The high resolution C1s XPS spectra of the nanoplatelets were deconvoluted into six bands (see Experimental Section for details). The low oxygen content observed for pristine GnP (1.74 atomic percentage - at.%), increased to 5.23 % after functionalization with 4-aminophenol. In GnP-OH a clear band at 285.7 eV, not detected in the pristine GnP, corresponding to the C-OH groups (Figure 1d) was observed, thus confirming the grafting of phenolic groups to the nanoflakes. On the other hand, in the C1s spectrum of E-GnP (Figure 1f) the band located at ~ 286.6 eV corresponding to the C-O-C groups is higher than in GnP-OH, which is a clear evidence of the successful functionalization of GnP in one step by arylation with diazonium salts. Indeed, it can be clearly observed in the O1s spectrum of E-GnP (Figure S7) that the vast majority of oxygen moieties corresponds to single bonded C-O-C groups. [6a] Furthermore, it is noteworthy that the oxygen content is 6.29 %, which is higher to that observed in GnP-OH. This confirms that the optimized conditions for the diazonium reaction with 4-aminophenol are valid also for the dianilinic compound **3**. These observations are confirmed by the values of the at.% of each oxygen functional group summarized in Table S2 and the bands assigned in the O1s spectra of the graphene nanoplatelets (Figure S7). Upon functionalization either by 4-aminophenol (Figure 1c) or **3** (Figure 1e), a weak N1s peak was observed on the XPS survey spectra, being N/C ratio 0.04 for both GnP-OH and E-GnP. Analysis of the N1s spectra (Figure S8) revealed one broad band in the range 398 – 400 eV, which can be ascribed to amine (-NH$_2$) and/or azo (-N=N-) groups. [16] Indeed, the presence of the NH$_2$ groups is attributed to the adsorption of the starting anilinic molecule, while the appearance of the azo groups can be related to the diazonium ion molecules generated in situ that form a charge-transfer complex with the graphene surface. [17]

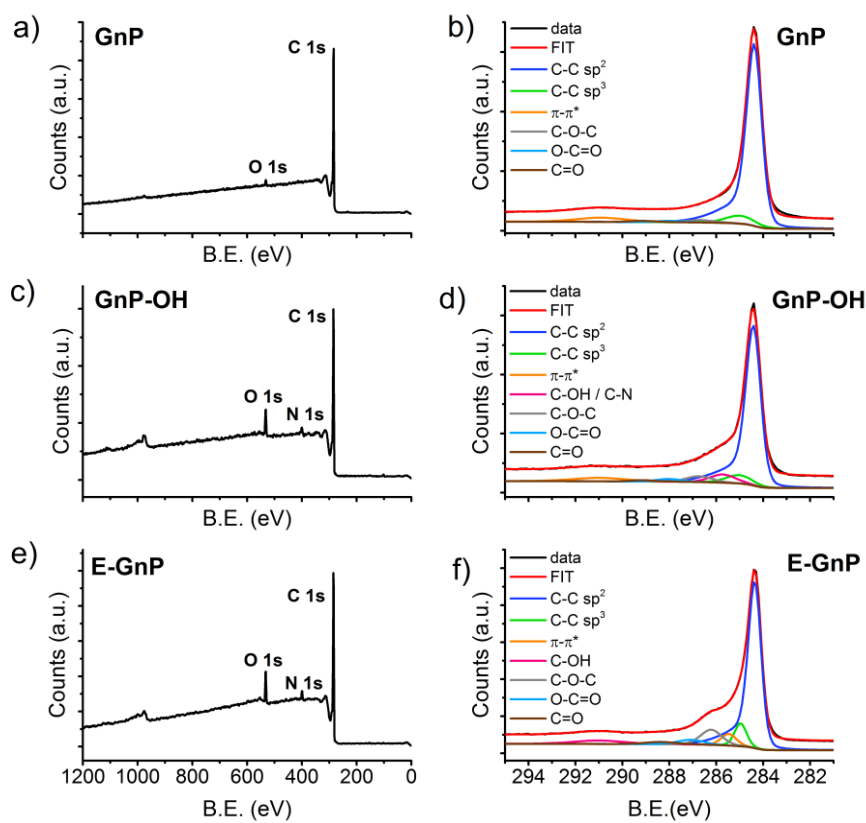

Figure 1. Survey XPS spectra and C1s XPS spectra of: GnP (a-b), GnP-OH (c-d), and E-GnP (e-f).

While the XPS analysis evidences the functionalization of GnP with –OH groups and the latter formation of the ether linkage, this technique provides no information on the location of the functional groups on the GnP flakes. In particular, despite the diazonium functionalization was previously shown to start from the edges of graphene sheets, [18] where the most reactive sites are located, the reaction occurring at the edges of GnP was also investigated.

Raman spectroscopy, because of its sensitivity to changes in atomic structures, has proved to be an appropriate technique to characterize the presence of disorder in *sp²* hybridized carbon-based systems. [19] In particular, it can determine the nature of disorder, from point defects and boundaries to zig-zag and armchair edges. In this study, Raman spectroscopy was used to obtain evidence of the preferential functionalization of graphene nanoflakes at the edges. The most characteristic peaks in the Raman spectrum of graphene-based materials are the so-called G band (~ 1575 cm$^{-1}$) associated with the doubly degenerate in-plane transverse/longitudinal optical phonon modes (i-TO/i-LO); the G' mode (~ 2700 cm$^{-1}$) due to a double resonance intervalley Raman scattering process with two in-plane transverse optical phonons (iTO) at the K point and the D band (~ 1350 cm$^{-1}$) arising from the iTO phonon mode near K points in the Brillouin zone. The D band is activated by structural defects by a second order Raman scattering process through the intervalley double resonance and thus its intensity is proportional to the amount of disorder in the sample. Thus, the ratio between the intensities of the D band and the G band ($I_D/I_G$) provides a parameter for quantifying disorder. [19b, 20] Spatial mapping of $I_D/I_G$ (Figure 2 a-c) of the different graphene nanoplatelets was useful to localize the regions with higher defect density. As expected, the $I_D/I_G$ ratio is lower in the basal planes of the unfunctionalized starting material than near the edges (Figure 2a). Upon functionalization to obtain GnP-OH and E-GnP the amount of disorder on the basal planes, observed in the corresponding spatial mappings, did not present major differences if compared to the unfunctionalized GnP, confirming that the grafting procedure does not introduce significant defectiveness in the *sp$^2$* structure. However, single-point Raman spectra recorded near the edges and the center regions of the different flakes (Figure 2 d-f) revealed an increase of the $I_D/I_G$ ratio at the borders.

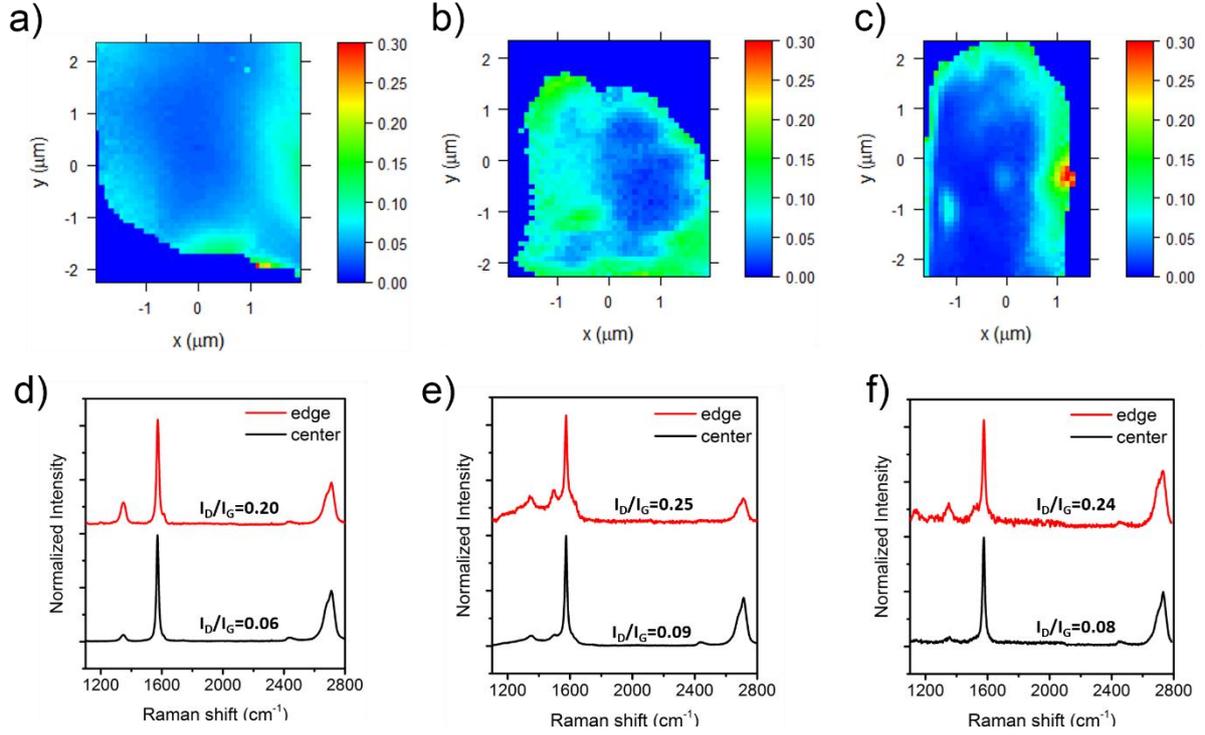

**Figure 2.** Raman mapping of $I_D/I_G$ of (a) GnP, (b) GnP-OH and (c) E-GnP. Average Raman spectra (d), (e) and (f) extracted from edge and center spots in mappings (a), (b) and (c).

The first-order Raman region (up to 2000 cm$^{-1}$) of points localized at the edges of the flakes were well fitted by Lorentzian functions (**Figure 3**). The unfunctionalized GnP shows the characteristic D and G bands, while the second disorder-induced peak around 1614 cm$^{-1}$ (the D' band) also becomes evident (Figure 3a). In GnP-OH, additional features at 1167, 1210, 1279, 1420, 1495 and 1542 cm$^{-1}$ were observed in the first-order Raman spectral region (Figure 3b). The peak at 1495 cm$^{-1}$ can be assigned to the iTO phonon branch, that becomes Raman active by the intra-valley double resonance Raman process. [19a, 21] The peak at 1167 cm$^{-1}$ coupled with the peak at 1420 cm$^{-1}$ are a signature of *trans*-polyacetylene, as reported by Ferrari *et al*. [22] On the other hand, the peak at 1210 cm$^{-1}$ and its companion feature at 1542 cm$^{-1}$ have been assigned to *cis*-polyacetylene. [23] *Trans*- and *cis*-polyacetylene present zigzag and armchair structures similar to the edges of graphene sheets. [24] Then the origin of these two pair of coupled modes, 1167 – 1420 cm$^{-1}$ and 1210 – 1542 cm$^{-1}$, was ascribed to C-C and C=C stretching bond vibrations of the edge atoms of graphene flakes, respectively, justified by interrupted conjugation at the edges of graphene sheets. Such interruption of conjugation in the graphene may be obtained as a consequence of grafting at the graphene edges. Clearly, in the reaction condition used, distance between two grafted molecules cannot be controlled and is expected to result in a wide distribution of conjugation length, which appear confirmed by the broad signals observed in the Raman spectra. Similar features were previously reported for edge-carboxylated nanosheets [25] and functionalized graphene with diazonium salts, [26] thus confirming the functionalization at the edges of the flakes. Finally, the observation of the C-O stretching vibrations for *p*-monosubstituted phenol at around 1279 cm$^{-1}$, [27] provides irrefutable evidence of the grafting of phenol groups at the edges of the GnP.

The Raman spectra of E-GnP does not clearly show the presence of the weak coupled bands previously observed in GnP-OH ascribed to interrupted conjugations (armchair and zig-zag structures) of the edges of graphene. Instead, features at 1140 and 1517 cm$^{-1}$ can be clearly observed in the Raman spectrum of E-GnP. Both bands are also observed in the Raman spectrum of the dianilinic compound **3** (Figure S9), thus further proving grafting in E-GnP. [27a] Furthermore, the Raman mapping of the ratio between the $I_{1140\ cm^{-1}}/I_G$, shown in Figure S10, confirms the preferential location of the alkyl chains at the edges of the graphene nanoplatelets. In addition, it is worth noting that the bands in the 1350 – 1500 cm$^{-1}$ region observed in the anilinic compound, ascribed to C-N bonds, are not detectable in the Raman spectrum of E-GnP, [27a, 28] further confirming the covalent grafting of **3**.

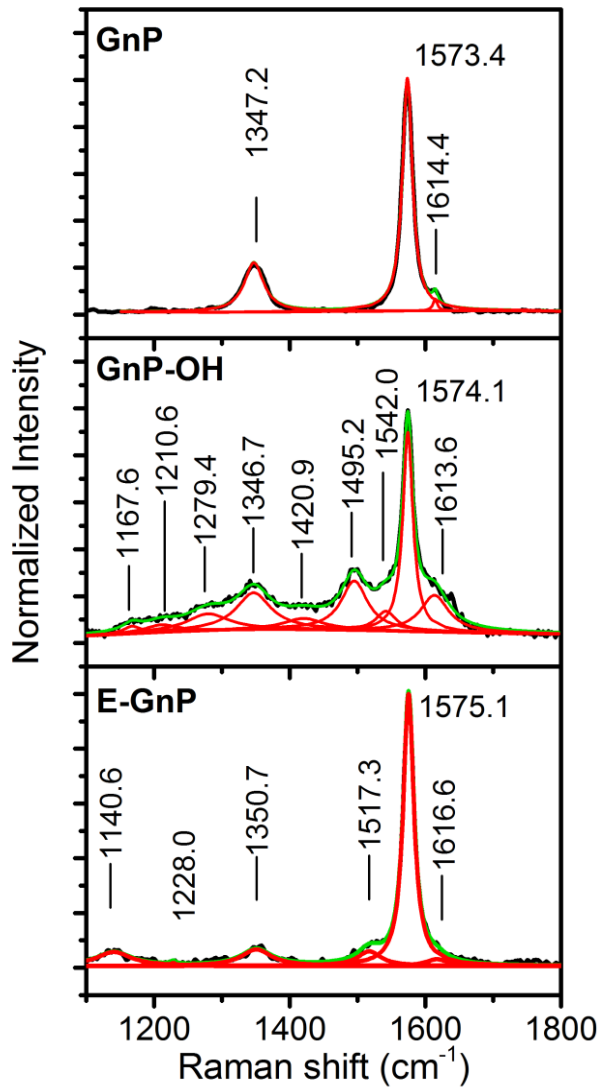

**Figure 3.** First-order Raman spectra recorded from the edges of GnP, GnP-OH and E-GnP with their deconvolution peaks.

To study the effect of chemical functionalization on thermal properties of junctions between GnP flakes, Molecular Dynamics (MD) calculations were carried out on a simplified model system, made of two adjacent graphene sheets edge functionalized with phenols or with the covalent molecular junctions (**Figure 4**), using a well-known method previously reported. [12c]

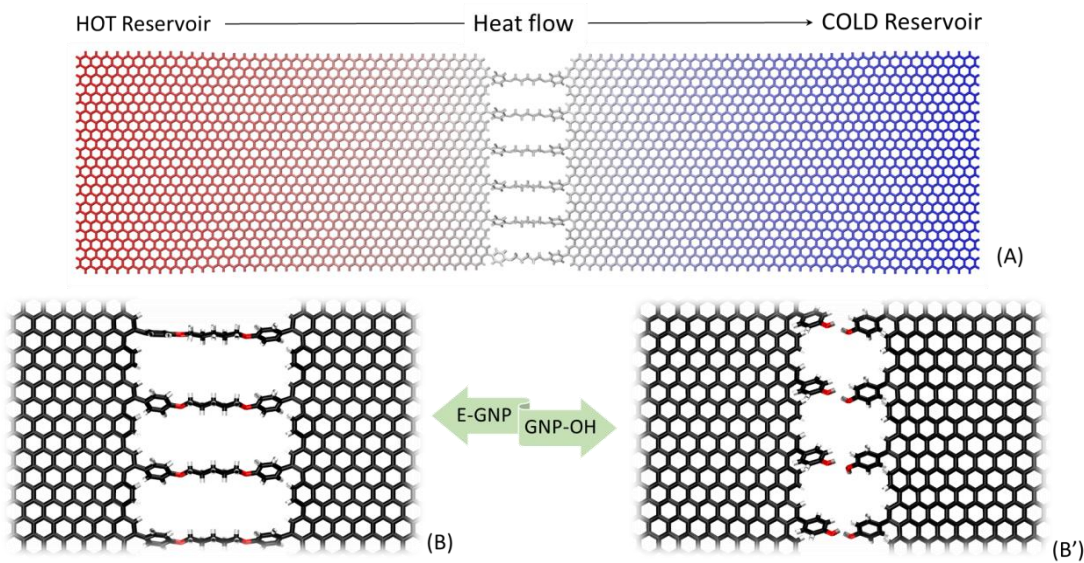

**Figure 4.** Atomistic models for the calculation of thermal conductance. A) full size model model for E-GnP, B and B') Magnifiction of representative screenshots from MD simulations for E-GnP and GnP-OH, respectively.

This method allowed to calculate a 126 pW/K thermal conductance per single chain in E-GnP, whereas a maximum conductance value of 22 pW/K was obtained per each couple of edge grafted interacting phenols, when minimizing the distance between hydroxyl groups. A detailed analysis of the thermal conductance dependence on the distance between nanoribbons is reported in Supplementary Information (Figures S13, S14). The factor 6 between the conductances for covalent *vs.* non-covalent junctions evidences significant differences on the different functionalization strategies, with a clearly higher thermal transfer efficiency in the presence of covalently bound molecular junctions. Beside the values calculated in idealized junctions, the occurrence of strong coupling between edges of GnP flakes is expected to be beneficial for thermal transfer at the interface in GNP networks, providing additional channels for heat transfer on top of the obvious overlapping of un-functionalized flakes. As the thermal conductance between parallel overlapping graphene sheets was previously calculated by MD to be 0.38 pW/Å$^2$K, [12c] it is possible to estimate the heat transfer equivalence of a single covalently bound molecular junction to the conductance obtained when overlapping of two graphene sheets for about 330 Å$^2$, which corresponds to about 130 carbon atoms per layer. In order to investigate the effect of edge functionalization, graphene nanopapers were prepared by vacuum filtration of suspensions of GnP and functionalized GnP. The free-standing nanopapers were obtained after peeling off from the membrane filter, drying under vacuum and mechanical pressing. The Field Emission Scanning Electron Microscopy (FESEM) images of the different nanopapers (**Figure 5**) reveal an aligned layered structure through the entire cross-section. As previously reported, [29] the vacuum filtration process produces a degree of order within the structure of the nanopapers, while the morphology of the nanoflakes strongly influences their self-assembly. In our case, the planarity of the starting GnP together with their low defectiveness allows the formation of an ordered and aligned layered structure (Figure 5a and 5b). However, the density of the nanopaper is about 1.3 g/cm$^3$, yielding a porosity of about 41%. In GnP-OH the ordered structure is retained (Figure 5c and 5d) which is also confirmed by the unmodified density of the nanopaper (Table S3). E-GnP nanopaper is not as highly aligned as GnP-OH nanopaper and yields a porosity of about 48 %. This effect is indeed ascribed to the presence of covalent junctions between different nanoflakes reducing their mobility to organize in a parallel way during filtration.

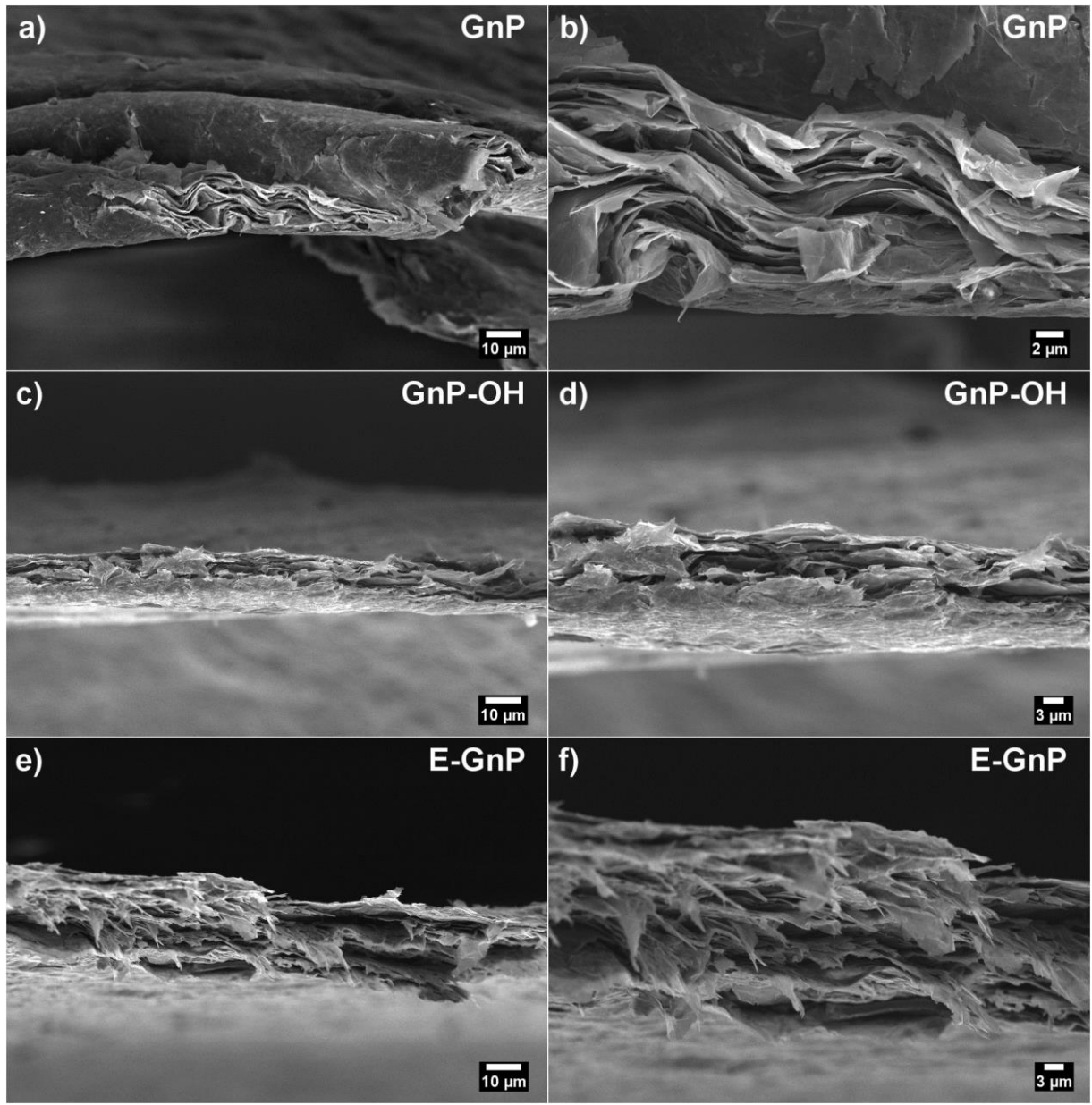

**Figure 5.** Cross-sectional FESEM image of (a) GnP, (c) GnP-OH, and (e) E-GnP and corresponding higher magnifications (b), (d), and (f), respectively.

The thermal conductivity of the graphene nanopapers was calculated from the values of the in-plane ($\alpha_\parallel$) and cross-plane ($\alpha_\perp$) thermal diffusivities obtained from the light flash measurements (Table 1). To take into account the differences in porosity between the different nanopapers, the effect of air on the conductivity of the nanopaper was subtracted using the well-known Maxwell's effective medium approach to calculate the effective conductivity of the GnP network, $k_n$ (see Experimental section for details). The values of the in-plane effective network conductivity ($k_{n\parallel}$) were found higher in both GnP-OH (266 W m$^{-1}$ K$^{-1}$) and E-GnP nanopapers (273 W m$^{-1}$ K$^{-1}$), compared to pristine GnP (228 W m$^{-1}$ K$^{-1}$). The increase in thermal conductivity for GnP-OH nanopaper confirms that a strong interaction between the nanoparticles is taking place, explained by the formation of hydrogen bonding. In the case of E-GnP nanopapers, covalent molecular junctions produced between nanoflakes account for a similar value of $k_{n\parallel}$ to that observed in GnP-OH, evidencing the effectiveness of the organic functionalization to mediate thermal transfer between nanoflakes. As expected, the values of the $k_{n\perp}$ are significantly lower compared to those of the $k_{n\parallel}$, thus indicating the high anisotropy of the graphene nanopapers. Beside the absolute values, it is worth noting the $k_{n\perp}$ for functionalized GnP follows a totally different trend respect to the $k_{n\parallel}$. Indeed, the $k_{n\perp}$ is reduced by 35 % in GnP-OH nanopaper, while $k_{n\perp}$ of E-GnP nanopapers is 190 % higher compared to pristine GnP nanopaper. Such dramatic differences are ascribed to the interactions driving self-assembly and orientation of the nanoflakes in the nanopapers. Indeed, while GnP-OH organizes in highly aligned nanopaper perpendicular to the direction of the filtration flow, thus favouring $k_{n\parallel}$, the presence of grafted phenols may hinder to some extent the vertical stacking of nanoflakes, thus reducing the efficiency of heat transfer in the through-plane direction. On the other hand, in the presence of covalent molecular junctions, the lower orientation observed contributes in increasing the through plane conduction. Besides the obvious effect of orientation, these results suggest an interplay between the formation of molecular junctions and GnP stacking, *i.e.* between conduction though molecular bridges *vs.* by π-π overlapping.

**Table 1.** In-plane ($\alpha_{\parallel}$) and cross-plane ($\alpha_{\perp}$) thermal diffusivities and in-plane ($k_{n\parallel}$) and cross-plane ($k_{n\perp}$) effective thermal conductivities of graphene nanopapers.

| Nanopaper | $\alpha_\parallel$ [mm² s⁻¹] | $\alpha_\perp$ [mm² s⁻¹] | $k_{n\parallel}$ [W m⁻¹ K⁻¹] | $k_{n\perp}$ [W m⁻¹ K⁻¹] |
|---|---|---|---|---|
| GnP | 121.4 ± 1.1 | 0.14 ± 0.01 | 228.4 ± 2.5 | 0.231 ± 0.001 |
| GnP-OH | 141.4 ± 4.5 | 0.10 ± 0.01 | 265.8 ± 9.8 | 0.151 ± 0.001 |
| E-GnP | 141.0 ± 2.5 | 0.37 ± 0.01 | 273.1 ± 5.8 | 0.672 ± 0.001 |

In order to compute the cumulative effect of $\pi$-$\pi$ interactions and molecular junctions into a value of thermal conductance, heat transfer was simulated by finite elements method on model nanopapers designed to match both nanoflakes size distribution and nanopaper density as detailed in the Methods section. As a result of the applied heat flux, a steady-state temperature profile establishes along the constructed sample (Figure 6).

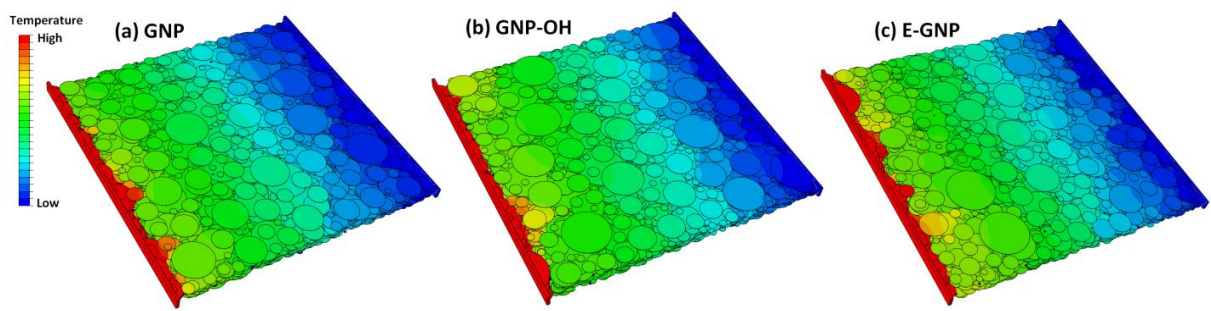

**Figure 6.** In plane temperature profile on nanopapers based on GnP, GnP-OH and E-GnP, from finite element analysis.

Based on our modelling results, the interfacial thermal conductance was calculated to be 17 MW/m²K, 19 MW/m²K and 26 MW/m²K for the GnP, GnP-OH and E-GnP graphene laminate samples, respectively. This value accounts for both heat transfer of $\pi$-$\pi$ interactions and molecular junctions, thus providing a cumulative evaluation of the efficiency of thermal contact between the nanoparticles. The results obtained confirm that both covalent and non-covalent molecular junctions are indeed beneficial in improving thermal contacts compared to pristine GnP, with a clearly higher success of covalently bound junctions. The higher efficiency of the covalent molecular junction was finally confirmed by a simple proof-of-concept heat spreader demonstrator (see Supporting Information), especially in the through-plane direction (Figure S16b), corroborating the results obtained from the LFA technique.

## 3. Conclusions

The functionalization of graphite nanoplatelets was obtained via diazonium reaction with either aminophenol or a bifunctional dianilinic compound synthesized on purpose. In both cases, successful covalent grafting was confirmed by XPS, particularly by the strong increase of C-OH and C-O-C signals, respectively. Furthermore, Raman spectroscopy and mapping allowed to prove the preferential location of functional groups at the edges of nanoflakes, allowing to obtain nanoplatelets which are edge-decorated with chemical functions suitable for the controlled self-organization into advanced nanomaterials. In particular, the edge functionalization obtained were found suitable to produce molecular junctions between nanoflakes for the enhancement of their thermal boundary conductance. Molecular dynamics investigation suggested a 6-fold higher conductance of the covalent molecular junctions designed in this work, compared to the secondary interaction between edge grafted phenolic groups. Beside the theoretical calculations, experimental evidences of the enhancement in heat transfer were obtained from the thermal characterization of nanopapers prepared by vacuum assisted filtration using the two different functionalized GnP, compared to nanopapers made of pristine GnP. In particular, GnP-OH were used to produce non covalent molecular junctions between nanoflakes, driven by hydrogen bonding between phenolic groups, while covalent molecular junctions were achieved using E-GnP. Both GnP-OH and E-GnP were demonstrated to enhance the in-plane thermal conductivity by about 20%, while the cross-plane thermal conductivity was dramatically enhanced by 150% in the case of E-GnP. These noticeable enhancements are clearly driven by the GnP functionalization, affecting both the self-assembly and orientation of the nanoflakes in the nanopapers. Indeed, while GnP-OH organizes in highly aligned nanopaper perpendicular to the direction of the filtration flow, the presence of covalent molecular junctions reduced nanoflakes orientation, contributing to the increase of the through-plane heat transfer. The overall enhancement of nanopaper conductivity is therefore associated with an interplay of heat transfer onto molecular junctions and secondary effects related to the different organization of nanoflakes, in terms of orientation and π-π stacking. The chemistry-controlled organization of nanoparticles was therefore validated as a fascinating route for the design and manufacturing of efficient nanomaterials for heat management, including applications as heat spreaders in electronics, as well as in low temperature heat exchange and heat recovery, currently of utmost interest for the energy efficiency of both industrial and household systems.

## 4. Experimental Section

All chemical reagents were purchased from Aldrich, Alfa Aesar, Fisher or Acros Organics and used without further purification unless otherwise stated. Graphene nanoplatelets (GnP, G2Nan grade, lateral size ~ 10-50 µm and flake thickness ~ 10 nm, 99.9%) were kindly supplied by Nanesa (I).

*Edge-selective functionalization of GnP with 4-aminophenol (GnP-OH)*

GnP were firstly dispersed in DMF at a concentration of 1 mg mL$^{-1}$ and sonicated in an ultrasonic bath (maximum operating power 320 W) for 30 min while $N_2$ is bubbled in the suspension. Then, 4-aminophenol was added to the dispersion and the mixture was sonicated for further 15 min under $N_2$ to achieve a homogeneous suspension. After that, isopentyl nitrite (6 equiv. per carbon atom) was slowly added to the dispersion and the temperature of the reaction mixture was raised to 80 °C. The reaction was performed under different conditions (Scheme S1). After cooling to room temperature, the reaction was quenched into distilled water and filtered through a PTFE membrane (0.2 µm, Whatman). The filtered cake was redispersed in DMF, sonicated in an ultrasonication bath for 10 min and filtered through a PTFE membrane (0.2 μm). This sequence was repeated twice with DMF, distilled water, methanol and diethyl ether. The resulting solids were dried at 80 ºC for 24 h.

*Covalently linked E-GnP*

GnP were firstly dispersed in DMF at a concentration of 1 mg mL$^{-1}$ and sonicated in an ultrasonic bath (maximum operating power 320 W) for 30 min while N$_2$ is bubbled in the suspension. Then, **3** was added to the dispersion and the mixture was sonicated for further 15 min under N$_2$ to achieve a homogeneous suspension. After that, isopentyl nitrite (6 equiv. per carbon atom) was slowly added to the dispersion and the temperature of the reaction mixture was raised to 80 °C. After cooling to room temperature, the reaction was quenched into distilled water and filtered through a PTFE membrane (0.2 μm, Whatman). The filtered cake was re-dispersed in DMF, sonicated in an ultrasonication bath for 10 min and filtered through a PTFE membrane (0.2 μm). This sequence was repeated twice with DMF, distilled water, methanol and diethyl ether. The resulting solids were dried at 80 ºC for 24 h.

*Nanopaper preparation*

GnP and functionalized GnP were suspended in DMF at concentrations of 0.15 mg mL$^{-1}$ and the solutions were sonicated in pulsed mode (15 s on and 15 s off) for 15 min with power set at 30 % of the full output power (750 W) by using an ultrasonication probe (Sonics Vibracell VCX-750, Sonics & Materials Inc.) with a 13 mm diameter Ti-alloy tip. The suspensions were subjected to vacuum filtration using a Nylon Supported membrane (0.45 μm nominal pore size, diameter 47 mm, Whatman). After filtration, the as-obtained papers were peeled off from the membranes and dried at 65 °C under vacuum for 2 hours to completely remove the solvent. Then, the graphene nanopapers were mechanically pressed in a laboratory hydraulic press (Specac Atlas 15T) under a uniaxial compressive load of 5 kN for 10 min at 25 °C. The density ($\rho$) of the samples was calculated according to the formula $\rho = m/V$, where $m$ is the mass of the nanopaper, weighed using a microbalance (Sensitivity: < 0.1 µg) and $V$ is calculated from a well-defined disk film using the average thicknesses measured as described in the literature. [30]

*Characterization methods*

XPS were performed on a VersaProbe5000 Physical Electronics X-ray photoelectron spectrometer with a monochromatic Al source and a hemispherical analyser. Survey scans and high resolution spectra were recorded with a spot size of 100 µm. The samples were prepared by depositing the GnP powders onto adhesive tape and keeping them under vacuum for 15 hours prior to the measurement to remove adsorbed molecules. A Shirley background function was employed to adjust the background of the spectra. Atomic ratios (at.%) were calculated from experimental intensity ratios and normalized by atomic sensitivity factors (carbon 0.25, oxygen 0.66 and nitrogen 0.42). The C1s peak was fitted considering the contribution of C-C bond $sp^2$-like using an asymmetric peak (Doniach-Šunjić shape), [31] previously calculated on freshly cleaved HOPG (ZYH grade, Mikromasch®), obtained asymmetry index ($\alpha$) 0.115. The curve fitting was performed using a Gaussian (80%)-Lorentzian (20%) peak shape by minimizing the total square-error fit. The full width at half-maximum (FWHM) of each peak was maintained between 1.3-1.4 eV. The C1s spectra is deconvoluted into several peaks: C-C $sp^2$ with binding energy at 284.4 ± 0.1 eV, C-C $sp^3$ at 285.0 ± 0.1 eV, C-OH at 285.7 ± 0.1 eV, C-O-C at 286.6 ± 0.2 eV, O-C=O at 288.0 ± 0.1 eV, C=O at 289.0 ± 0.1 eV and $\pi$-$\pi$* shake-up satellite peak from the $sp^2$-hybridized C atoms at 291.0 ± 0.2 eV. [6a, 11, 32]

Raman spectroscopy measurements were performed exciting the GnP samples with a 514.5 nm laser coupled to a Renishaw inVia Reflex (Renishaw PLC, United Kingdom) microRaman spectrophotometer. A long working distance 100x objective was employed for the acquisition in backscattering configuration using a laser power of 2.5 mW and an integration time of 15 s. The spectral resolution was 3 cm$^{-1}$. Raman maps were collected after drop-deposition and drying of the GnP flakes on $SiO_2$/Si substrates with a 100 nm step over a grid including the selected flake area.

The morphology of the graphene papers was characterized by a high resolution Field Emission Scanning Electron Microscope (FESEM, ZEISS MERLIN 4248).

The in-plane thermal diffusivity ($\alpha_\parallel$) and cross-plane diffusivity ($\alpha_\perp$) were measured using the xenon light flash technique (LFT) (Netzsch LFA 467 *Hyperflash*). The samples were cut in disks of 23 mm with thicknesses between 10 – 30 µm and the measurement of the $\alpha_\parallel$ was carried out in a special in-plane sample holder while the $\alpha_\perp$ was measured in the standard cross-plane configuration. Each sample was measured five times at 25 °C.

The in-plane and cross-plane thermal conductivities of the nanopapers, $k_\parallel$ and $k_\perp$ respectively, were then calculated from the equation $k = \rho \alpha C_p$, where $\rho$ is the density of the graphene film and $C_p$ is the specific heat capacity of graphite ($C_p = 0.71$ J (g K)$^{-1}$). In order to properly take into account the differences and porosity between the different nanopapers, the effect of air was subtracted assuming the nanopaper as a composite in which the continuous matrix is made of GnP particles and the inclusion is air. On such a composite, the well-known Maxwell's effective medium approach was applied (both in-plane and cross-plane) to calculate the effective conductivity of the continuous phase, i.e. the network of nanoflakes, $k_n$, from equation 1, where $k$ is the thermal conductivity of the nanopaper and $k_{air}$ is the thermal conductivity of air.

$$k = k_n \frac{\kappa_{air} + 2k_n + 2\varphi(k_{air} - k_n)}{\kappa_{air} + 2k_n - \varphi(k_{air} - k_n)} \qquad (1)$$

*Computational methods*

Classical Molecular Dynamics (MD) calculations was carried out on LAMMPS (Large-scale Atomistic Molecular Massively Parallel Simulator) package code which implements Velocity Verlet as integration algorithm to recalculate positions and velocities of the atoms. The class

II COMPASS force field was adopted in this work, its functional forms are described by Sun.[33]

The model was composed by two graphene nanoribbons (about 100Å by 50 Å) connected through the armchair edge by grafted molecules as depicted in Figure 4. The secondary interaction between phenols in GnP-OH were defined by the sum of the VdW contribution from the built-in Lennard Jones 9-6 function and the electrostatic interaction, represented by atomic partial charges.[33] Qeq-equilibration [34] of atomic partial charges was set up in model design.

Grafting density was kept constant at one grafted molecule per couple of aromatic rings on the edge, yielding a total of 6 grafted molecules on the width of the graphene sheet. The equilibrium distance between the GnP-OH sheets was varied in the range 7.9-15.0 Å and eventually adjusted to 12.5 Å at which the distance between OH groups was minimum, *i.e.* energy of the secondary interaction was maximum. Details are reported in SI. Linear conformation of chains in E-GnP was considered, yielding a 19.6 Å equilibrium distance between graphene sheets. Fully periodic conditions (no replicates) were used along X (length), Y (width) and Z (height). NEMD calculations were carried out applying Nosé-Hoover thermostats at the ends of the simulation box, *i.e.* the 10 Å graphene sheets ends. The Hot (310K) and the cold bath (290K) of the thermostats regions were set as NVT canonical ensemble (constant Number of atoms, Volume and Temperature) while the region between the two thermostats were set under NVE (constant Number of atoms, Volume and Energy) condition.

All the simulations were carried out for 15 Million timesteps (0.25 fs/ts), with an initial 500Kts NVE equilibrium at 300K and 1Mts thermo-stated preheating, followed with the purpose to reach a constant heat flux. After those initial stages, the constant energy flowing through the thermostats started recording. The thermal flow inside NVE regions is calculated from the slope of energy versus time plots.

Temperature profile along the system was calculated by virtually splitting the simulation box transversally into 22 thermal layers. The temperature of each thermal layer was computed by Equation 2, where $T_i(\text{slab})$ is the temperature of $i^{\text{th}}$ slab, $N_i$ is the number of atoms in $i^{\text{th}}$ slab, $k_B$ is the Boltzmann's constant, $m_j$ and $p_j$ are atomic mass and momentum of atom $j$, respectively.

$$T_i(slab) = \frac{2}{3N_i k_B} \sum_j \frac{p_j^2}{2m_j} \qquad (2)$$

Temperatures were time-averaged to the simulation runtime excluding the non-linear regions at the interfaces, i.e. both close to the thermostats and across the junction.

Single chain thermal conductance $G_s$ expressed in pW/K has been calculated by Equation 3, where $q_x$ is the thermal flow derived from the energy versus time plot slope, 6 is the number of chains and $\Delta T$ is the temperature difference across the jump, as projection of the two linear fit of the temperature-length graph in the junction middle point.

$$Gs = \frac{q_x}{6 \cdot \Delta T} \qquad (3)$$

Finite element modeling in this study was conducted using the Abaqus/Standard (Version 6.14) package along with the python scripting. As a common assumption, individual GnP were modeled using the disc geometry. Moreover, in agreement with experimental samples we randomly distributed GnP in a way that exactly satisfies the experimentally measured size distributions and porosity of the nanopapers. In our modeling we constructed relatively large samples including over 4500 individual GnP flakes stacked in 25 layers up together. The thermal conductivity of multi-layer graphene was assumed to be 1300 W/mK, according to the experimental measurements by Ghosh *et al*. [35] We remind that in the graphene laminates the heat percolates not only through the particles but primarily through the contacting surfaces between individual particles. To simulate such a phenomenon, we introduced contact elements between every two contacting flakes with a constant interfacial thermal conductance. For the evaluation of effective thermal conductivity, we included two highly conductive strips at the two ends of the constructed samples which were thermally tied to the graphene flakes. [36] We then applied a constant inward and outward surface heat flux ($q$) on the external surfaces of the included strips. As a result of applied heat flux, a steady-state temperature profile establishes along the constructed sample. The established temperature difference along the laminate, $\Delta T$, was then used to acquire the effective thermal conductivity, $k_{eff}$, using one-dimensional form of the Fourier law (Equation 4):

$$k_{eff} = q \frac{L}{\Delta T} \tag{4}$$

Here, $q$ is the applied heat flux and $L$ is the laminate length (excluding the attached strips). After constructing the finite element models, we varied the interfacial thermal conductance between the graphene flakes to match the modeling results for the effective thermal conductivity with experimental measurements.

**Acknowledgements**

This work has received funding from the European Research Council (ERC) under the European Union's Horizon 2020 research and innovation programme grant agreement 639495 — INTHERM — ERC-2014-STG. The authors gratefully acknowledge Francesco Bertocchi at Nanesa (I) for providing GnP, Dr. Mauro Raimondo and Salvatore Guastella at Politecnico di Torino-DISAT for FESEM and XPS analyses, respectively. Fabio Carniato at Università del Piemonte Orientale (I) is gratefully acknowledged for NMR measurements as well as Fausto Franchini at Politecnico di Torino-DISAT for his precious assistance in the heat spreader demonstrator setup and measurements.

# Supporting Information

**Optimization of the edge-selective functionalization of GnP via in situ generation of diazonium species**

The optimization of the reaction conditions for the edge-selective functionalization of our GnP, using the diazonium chemistry, was initially investigated. We performed three different functionalizations of GnP by using 2 and 4 equiv. of the selected aniline, in this case 4-aminophenol, per C atom and two different reaction times, 24 h and 48 h (Scheme S1).

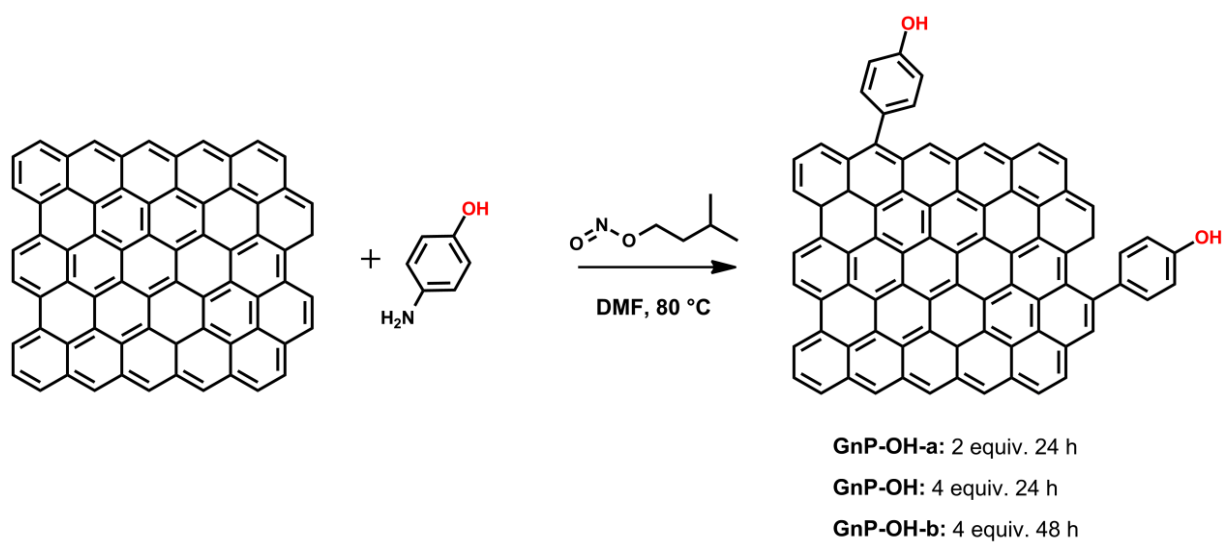

**Scheme S1.** Edge-selective functionalization of GnP with 4-aminophenol: 2equiv. 24a h (GnP-OH-a), 4 equiv. 24 h (GnP-OH) and 4 equiv. 48 h (GnP-OH-b).

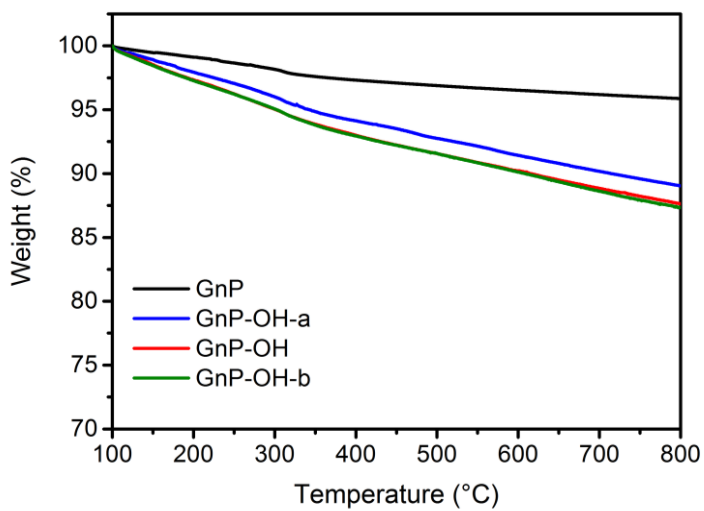

**Figure S1.** TGA of GnP, GnP-OH-a, GnP-OH and GnP-OH-b under nitrogen atmosphere (heating rate 10°C min$^{-1}$).

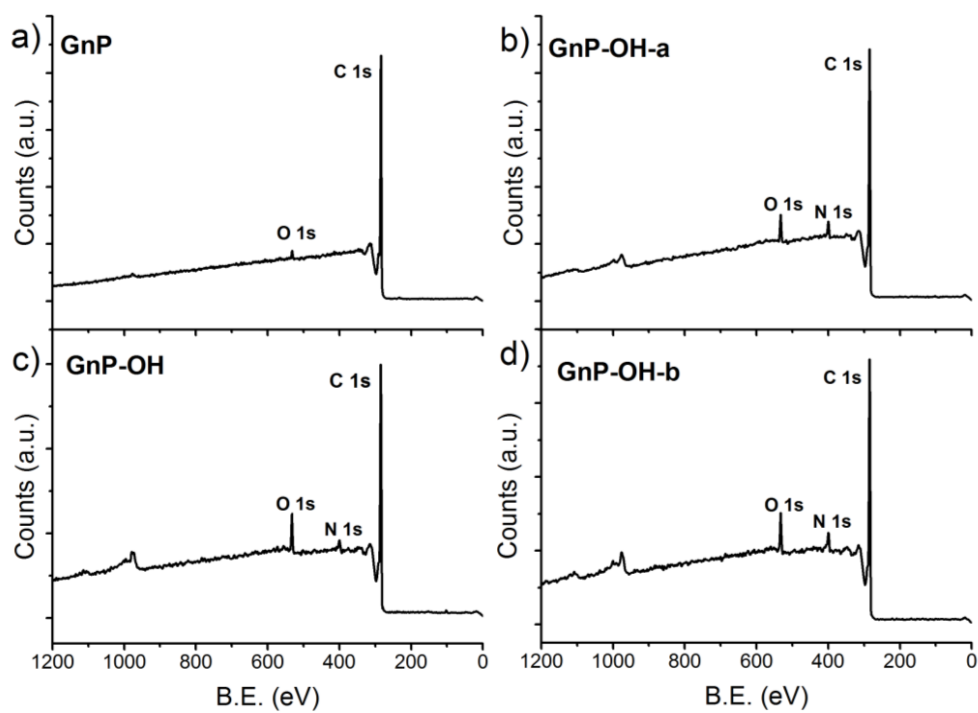

**Figure S2.** Survey XPS spectra of: (a) GnP, (b) GnP-OH-a, (c) GnP-OH and (d) GnP-OH-b.

The XPS analysis evidenced an increase of oxygen content (at.%) from 1.74 in GnP to a maximum of 6.29 in GnP-OH-b (Table S1), where the number of equivalents of the anilinic molecule per C atom in the reaction is higher, which is a first indication of the introduction of oxygen-bearing groups on graphene structure. High resolution C1s core level spectra of GnP and functionalized GnP with 4-aminophenol is shown in Figure S3. Upon functionalization, both the C-C *sp²* content and the π-π* shake-up band decrease due to the disruption of the delocalized π conjugation in the graphitic structure, while it is observed an increase of the relative intensity of the band corresponding to the C-OH groups, mainly in GnP-OH and GnP-OH-b prepared with higher contents of 4-aminophenol.

The contribution of the oxygen groups to the C1s spectra were estimated by calculating first the area percent of each corresponding oxygenated group and then the at.% were obtained using the O/C ratios for the individual functional groups (i.e. 1:1 for C=O).[1] results are summarized in Table S1. The amount of C-OH groups increases with the equivalents of aminophenol, while decreases with the reaction time. Indeed, the highest values of C-OH groups were observed for the reaction carried out with 4 equiv. of anilinic compound per C atom for 24 hours. The decrease on the at.% of C-OH groups for higher reaction times may be ascribed to the formation of other species as previously reported in similar systems.[2]

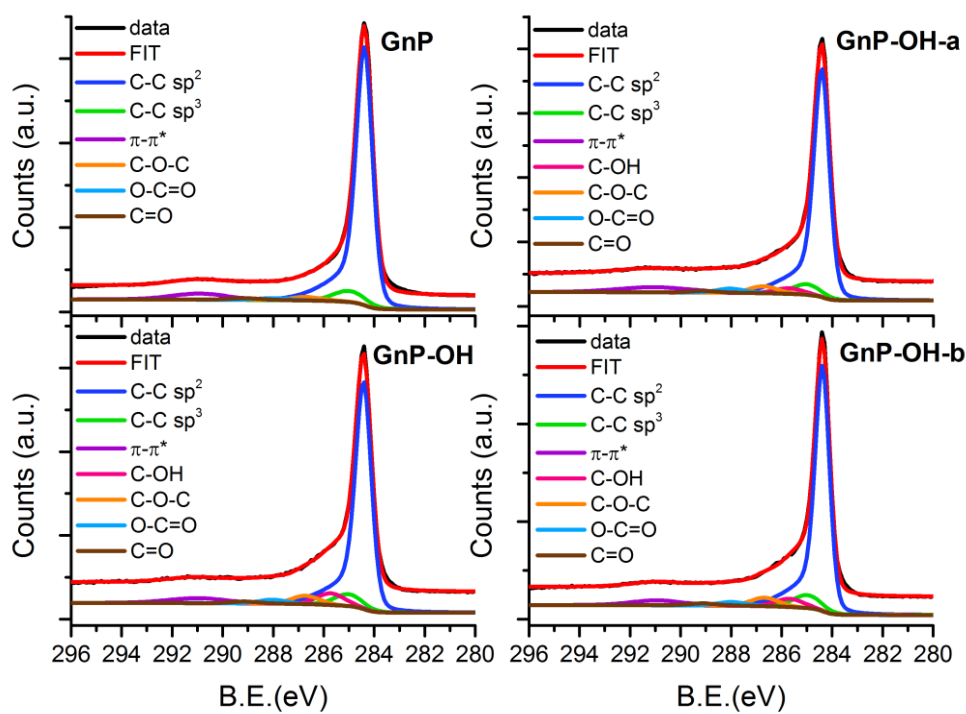

**Figure S3.** XPS spectra with their deconvolution peaks for C1s of GnP (a), GnP-OH-a (b), GnP-OH (c) and GnP-OH-b (d).

**Table S1**. XPS peak assignment and atomic percentage (at.%) for GnP, GnP-OH-a, GnP-OH and GnP-OH-b.

| Sample | Total C1s at.% | Total O1s at.% | Total N1s at.% | Contributions from O to C1s spectra at.% | | | |
|---|---|---|---|---|---|---|---|
| | | | | C-OH | C-O-C | O-C=O | C=O |
| **GnP** | 98.26 | 1.74 | - | 0.00 | 3.25 | 0.82 | 0.86 |
| **GnP-OH-a** | 91.86 | 4.92 | 3.22 | 2.46 | 4.60 | 0.92 | 0.78 |
| **GnP-OH-** | 91.48 | 5.23 | 3.29 | **4.85** | 5.19 | 0.81 | 0.82 |
| **GnP-OH-b** | 87.48 | 6.29 | 6.21 | 3.38 | 5.24 | 0.79 | 1.20 |

The O1s spectrum is deconvoluted into four bands: C=O at 530.5 ± 0.2 eV, O-C=O at 531.5 ± 0.2 eV, C-O-C at 533.0 ± 0.3 eV and C-OH at 534.2 ± 0.3 eV. The O1s spectra confirm the presence of the functional groups obtained from the C1s spectra as observed in Figure S4. While C-OH groups are not found in GnP, the same are clearly observed in functionalized graphene nanoplatelets, confirming the successful functionalization.

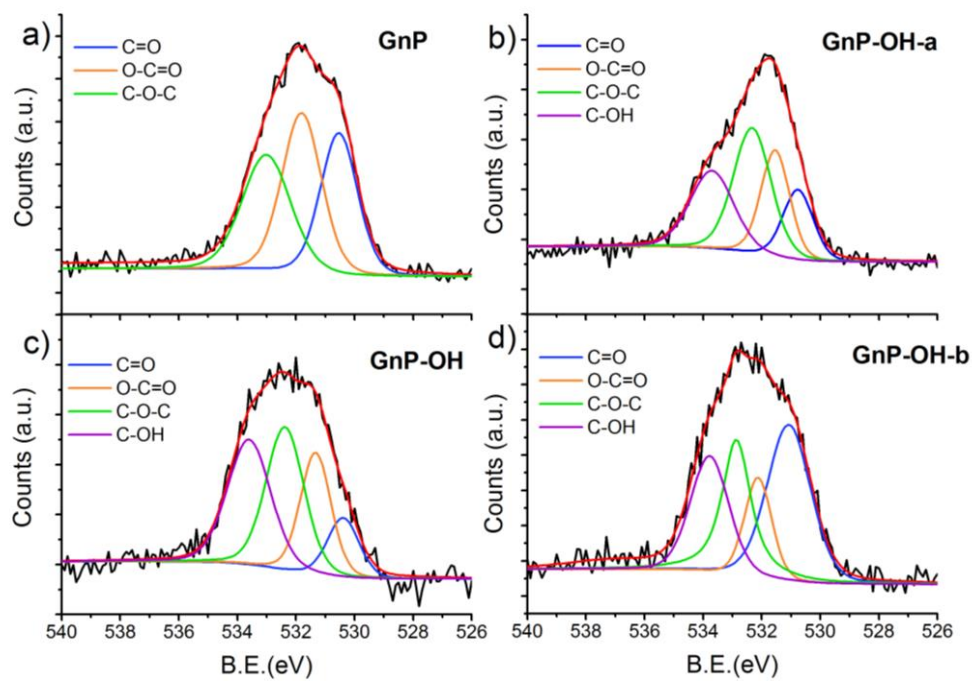

**Figure S4.** XPS spectra with their deconvolution peaks for O1s of GnP (a), GnP-OH-a (b), GnP-OH (c) and GnP-OH-b (d).

As described in the full text, weak peaks corresponding to N1s were observed in the survey spectra of functionalized graphene nanoplatelets, which are ascribed to –$NH_2$ and –N=N- groups, which are residual from the excess of anilinic compound and the diazonium chemistry reaction.

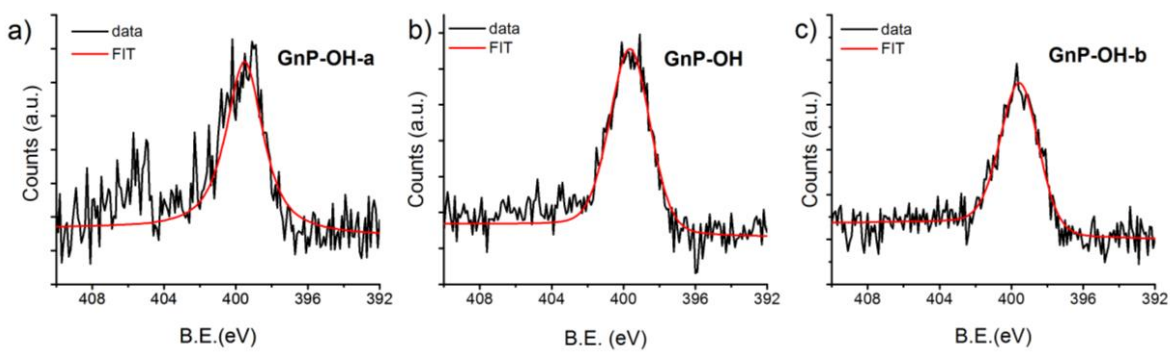

**Figure S5.** XPS spectra with their deconvolution peaks for N1s of GnP-OH-a (a), GnP-OH (b) and GnP-OH-b (c).

The Raman spectra of GnP-OH-a, GnP-OH and GnP-OH-b are shown in Figure S6a. After functionalization with 4-aminophenol it can be observed an increase of the D and D' band, as well as the appearance of the new bands at 1167, 1210, 1279, 1420, 1495 and 1542 cm$^{-1}$ (see full text). The highest values of $I_D/I_G$ were obtained for the samples GnP-OH and GnP-OH-b where higher amounts of the anilinic compound were used, corroborating the results observed in XPS and TGA.

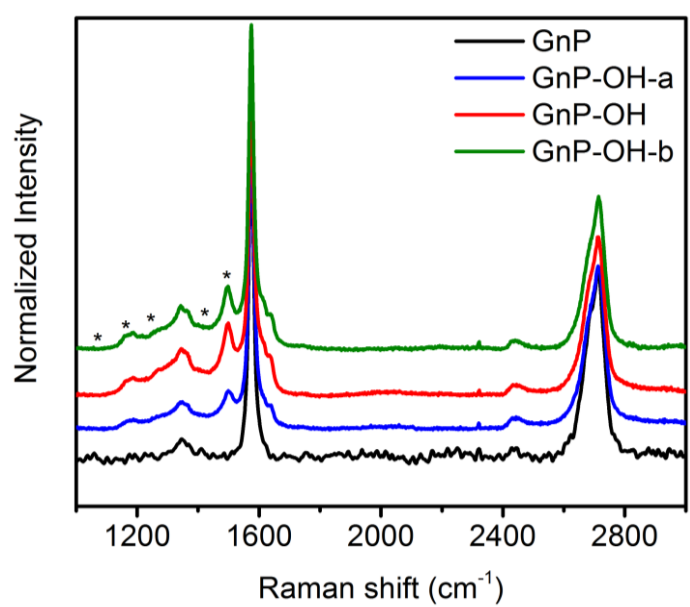

**Figure S6.** Raman spectra of GnP and functionalized GnP, GnP-OH-a, GnP-OH and GnP-OH-b**.** The new bands are marked with *.

According to the previous results, the best conditions for the edge-selective functionalization of graphene nanoplatelets using diazonium chemistry were identified with 4 equiv. of an anilinic molecule per C atom and 24 hours of reaction time.

**Synthesis of dianiline 3.**

All ¹H and ¹³C Nuclear Magnetic Resonance (NMR) spectra were recorded on a NMR 500 MHz Bruker AVANCE III. Samples were dissolved in deuterated dimethylsulfoxide (DMSO-$d_6$) with TMS as internal reference (chemical shifts δ in ppm). The following abbreviations were used to describe spin multiplicity: s = singlet, d = doublet, t = triplet, m = multiplet.

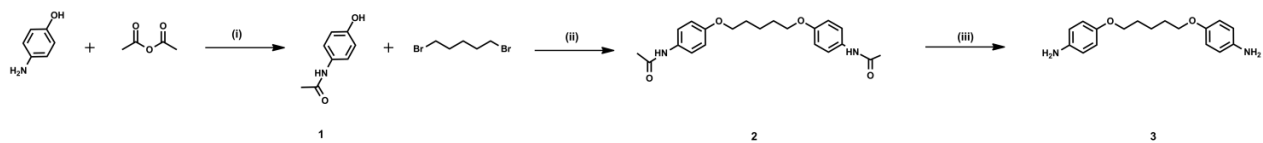

**Synthesis of N-(4-hydroxyphenyl)acetamide (1).**

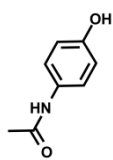

To a solution of 4-aminophenol (8.0 g, 73.3 mmol) in 120 mL of ethanol, acetic anhydride (6.9 mL, 73.3 mmol) was added. The solution was stir for 30 min at room temperature, and then evaporated to dryness. Purification of the solid by silica column chromatography (eluent dichloromethane/methanol 95:5) gave **1** as a white powder (8.8 g, 79.5 %). ¹H NMR (500 MHz, DMSO-$d_6$, 300 K) δ (ppm): 2.04 (s, 3H), 6.73 (d, 2H), 7.39 (d, 2H), 9.21 (s, 1H), 9.71 (s, 1H). ¹³C NMR (126 MHz, DMSO-$d_6$, 300 K) δ (ppm): 24.19, 115.45, 121.27, 131.50, 153.58, 167.95.

**Synthesis of 1,5-bis(4-acetamidophenyloxy)pentane (2).**

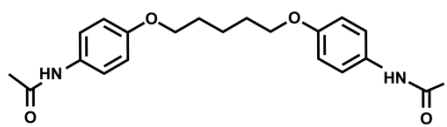

To a solution of **2** (8 g, 52.6 mmol) in 120 mL of acetone, potassium carbonate (16.4 g, 118.6 mmol) and 1,5-dibromopentane (3.6 mL, 26.4 mmol) were added. The reaction mixture was heated to reflux for 24 h. The mixture was cooled and poured into distilled water (1.5 L). The solid residue was filtered and recrystallized from ethanol, to give a white powder (8.8 g, 45.7 %). ¹H NMR (500 MHz, DMSO-$d_6$, 300 K) δ (ppm): 1.59 (s, 2H), 1.80 (s, 2H), 2.05 (m, 6H), 3.41 (s, 2H), 3.98 (t, 4H), 6.91 (m, 4H), 7.51 (m, 4H), 9.82 (s, 2H). $^{13}$C NMR (126 MHz, DMSO-d$_6$, 300 K) δ (ppm): 22.74, 24.26, 28.97, 67.96, 114.86, 120.97, 132.93, 154.89, 168.13.

**Synthesis of 1,5-bis(4-aminophenyloxy)pentane (3).**

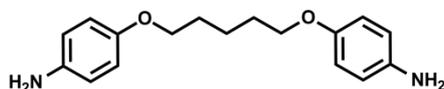

A mixture of **2** (8.0 g, 21.6 mmol), sodium hydroxide (36.7 g, 0.91 mol) dissolved in water (32 mL) and ethanol (32 mL) was heated to reflux overnight. The mixture was cooled and then evaporated to dryness. The resulting mixture was poured into ice-water (150 mL) and the pale brown precipitate was filtered and washed with distilled water. The solid was recrystallized from ethanol to afford **3** (3.24 g, 62.6 %). $^1$H NMR (500 MHz, DMSO-d$_6$, 300 K) δ (ppm): 1.57 (m, 2H), 1.75 (m, 4H), 3.88 (m, 4H); 4.63 (s, 4H), 6.55 (m, 4H); 6.69 (m, 4H). $^{13}$C NMR (126 MHz, DMSO-d$_6$, 300 K) δ (ppm): 22.82, 29.19, 68.37, 115.37, 115.71, 142.77, 150.48.

**XPS Characterization of functionalized GnP.**

**XPS O1s**

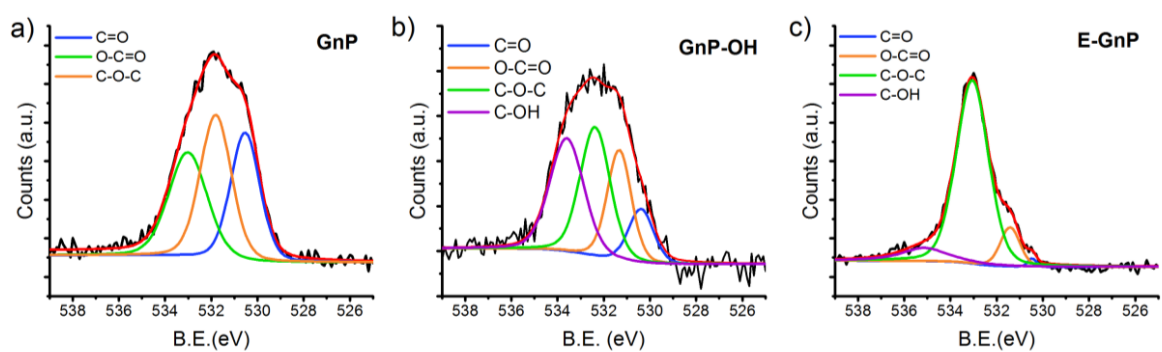

**Figure S7.** XPS spectra with their deconvolution peaks for O1s of a) GnP, b) GnP-OH and c) E-GnP.

## XPS N1s

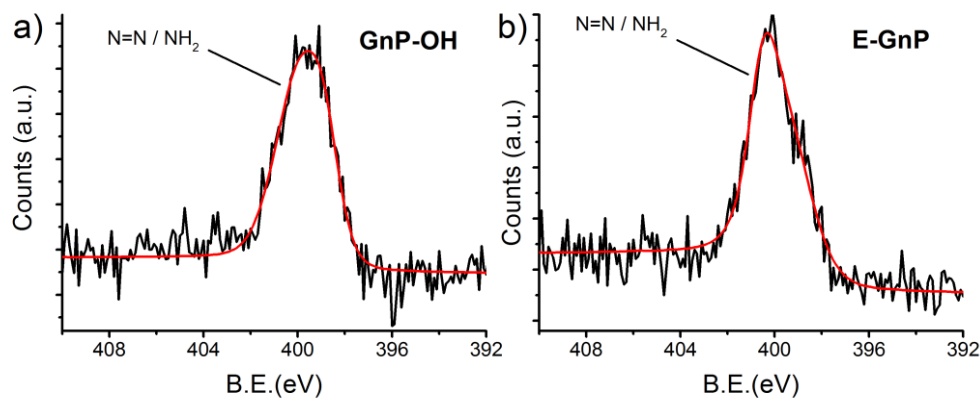

**Figure S8.** XPS spectra with their deconvolution peaks for N1s of a) GnP-OH and b) E-GnP.

**Table S2**. XPS peak assignment and atomic percentage (at.%) for GnP, GnP-OH and E-GnP.

| Sample | Total C1s at.% | Total O1s at.% | Total N1s at.% | Contributions from O to C1s spectra at.% | | | |
|---|---|---|---|---|---|---|---|
| | | | | C-OH | C-O-C | O-C=O | C=O |
| GnP | 98.26 | 1.74 | - | 0.00 | 3.25 | 0.49 | 0.86 |
| GnP-OH | 91.48 | 5.23 | 3.29 | **4.85** | 5.19 | 0.81 | 0.82 |
| E-GnP | 90.54 | 6.29 | 3.17 | 2.60 | **11.05** | 1.43 | 1.75 |

**Raman Characterization of functionalized GnP.**

Figure S9 shows the Raman spectra of 4-aminophenol and the dianilinic molecule **3**, synthesized in this study. The 4-aminophenol shows a clear band at 1260 cm$^{-1}$ corresponding to phenol groups and two other bands at 1165 and 1612 cm$^{-1}$ ascribed to C-O stretching and aromatic C=C stretching, respectively.[4] The Raman spectrum of **3** presents a clear band at 1139 cm$^{-1}$ which is assigned to C-C stretching from the alkyl chains, a broad band around 1520 cm$^{-1}$ and two bands at 1593 and 1611 cm$^{-1}$ corresponding to aromatic C=C stretching modes. The several bands located in the region between 1350 to 1500 cm$^{-1}$ are associated to C-N bonds.[3]

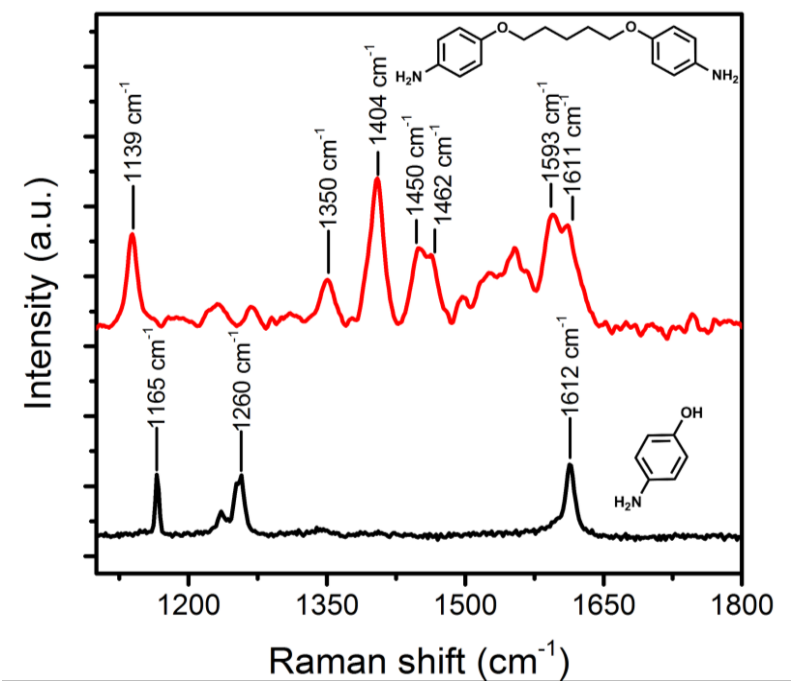

**Figure S9.** Raman spectra of 4-aminophenol (black) and **3** (red).

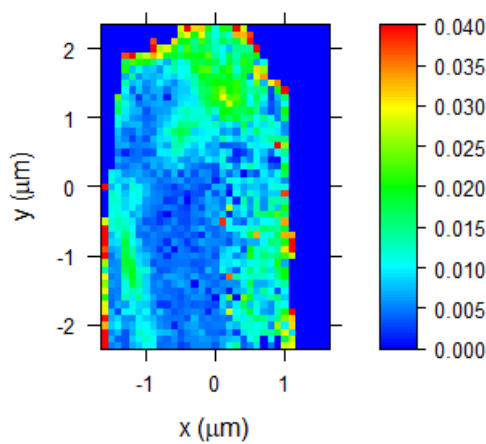

**Figure S10.** Raman mapping of $I_{1140\ cm^{-1}}/I_G$ of E-GnP.

**Molecular Dynamics for thermal conductance of GnP-OH and E-GnP**

Figure S11 represents the plot of Energy added to the hot reservoir and removed from the cold reservoir as a function of the time. The plots were linear and symmetric meaning that the total energy inside the ensemble is kept constant and steady heat flow passes through the model. A significantly higher slope is reached from the covalent-bonded scheme (approx. 0.09 versus 0.01 eV/ps).

The temperature profiles along the graphene nanoribbons and across the interface between those are reported in Figure S12. Herein, linear fitting of the temperatures along the interface allowed to calculate the temperature jump. For the covalent bonded nanoribbons we obtained 19.16 K thermal jump at the interface against 19.92 K calculated for the phenol functionalized nanoribbons.

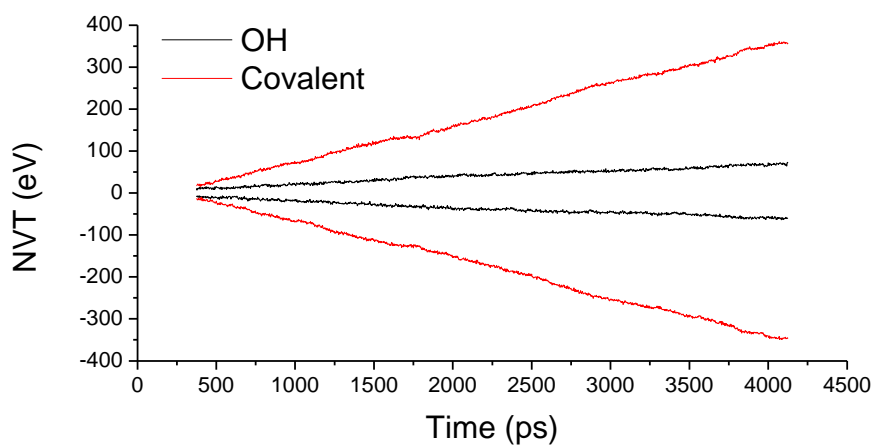

**Figure S11.** Energy added to the hot reservoir (positive slopes) and removed from the cold reservoir (negative slopes) for the covalent bonded model (in red) and the phenol system (black line).

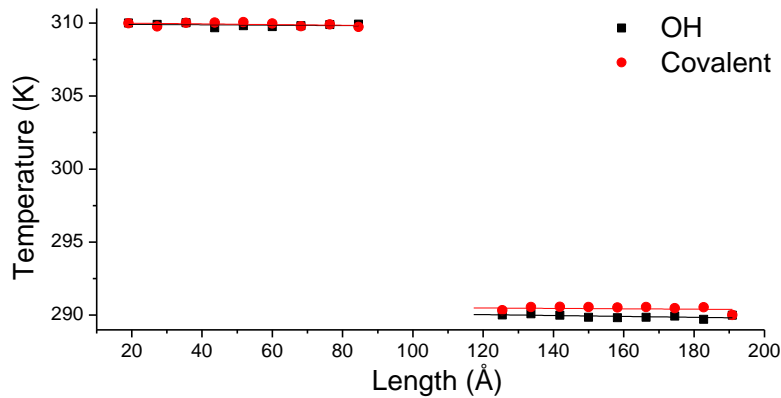

**Figure 12.** Temperature profile along the nanoribbons for covalent and non-covalent bonded nanoribbons. Dots represents the thermal layers while lines are linear fit.

Figure S13A represents thermal conductance as a function of the platelets distance, which clearly affects the distance between hydroxyls, thus the strength of the secondary interactions between the polar groups. Distances were routinely measured during simulation, using VMD software. A single direct measure (SDM) of distance between selected atoms was done along all the simulation time and considering the equilibrium value. The average of multiple SDM determined the plotted value as follows: for the measure of the platelets distance six SDM were involved; the final value is the average, where the maximum and the minimum values were discarded. This procedure helped to reduce the measure error from planarity and parallax. On the other hand, the measurements of the OH distance involved all the six phenols couples of OH groups (O left - H right; H left - O right) determining 12 SDM for every OH final value. In this case, all the SDM were considered in calculus.

For platelets distance below 8.5 Å, the thermal conductance exhibits values in the range of 20 ±2 pW/K), whilst the conductance rapidly decreases to about half the value increasing the platelets distance in the range of 9 to 10Å, despite limited changes in the average distance between OH groups were observed (Figure S13). By the analysis of the relative positions of the phenol groups during the simulation (Figure S14A), this phenomenon was attributed to $\pi - \pi$ interactions becoming dominant below 8.5 Å, where aromatic rings partially overlap. At slightly higher nanoribbon distance, aromatic ring overlapping was not observed and the decrease in the interface conductance reflect the weak VdW interaction between hydroxyls. When further increasing the nanoribbons distance, phenol groups are allowed to rearrange to minimize the distance between hydroxyls (Figure 14B), down to about 3.2 ±0.1 Å for nanoribbons distance in the range 12.2-12.8 Å (Figure 13B).The minimization of hydroxyl distance results in the maximization of interface conductance, to about 22 ±2 pW/K, exploiting to the strongest interactions between phenolic groups. A further increase of the platelets distance determines a rapid increase in the hydroxyl distances, corresponding to a dramatic decay in thermal conductance, towards non bonded nanoribbons.

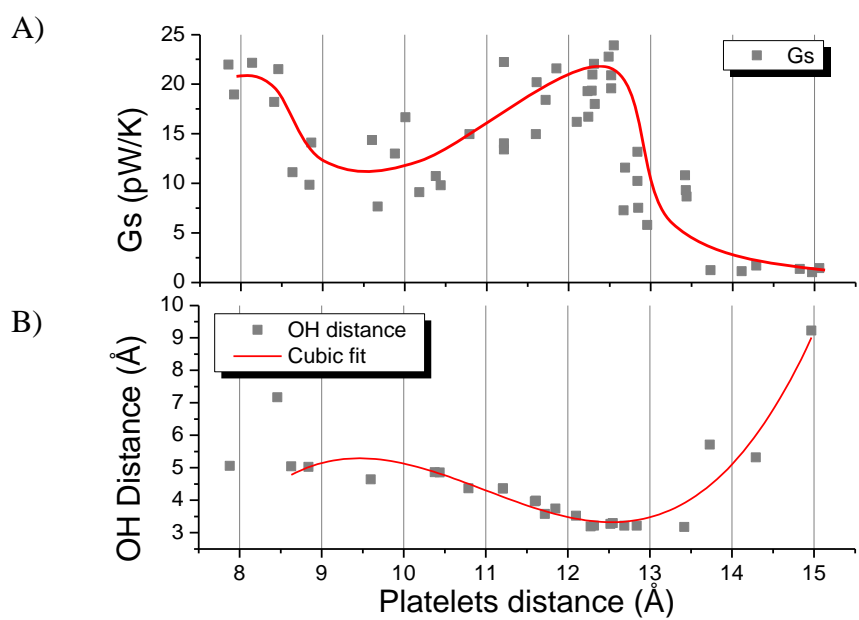

**Figure S13.** (A) Thermal conductances as a function of the platelets distances for the 7.8-15.3 Å region. The red line drives the eye through data (B) OH distance as a function of the distance between nanoribbons, red line is cubic fit of data.

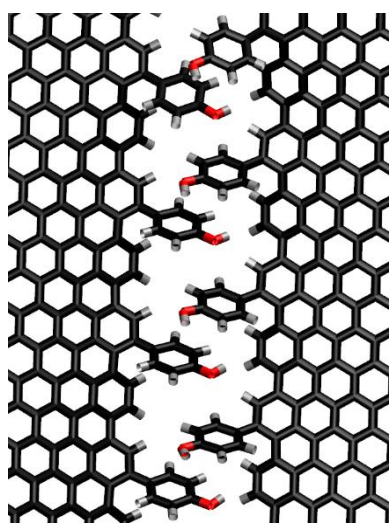

**(A)**     9.8 Å platelets distance

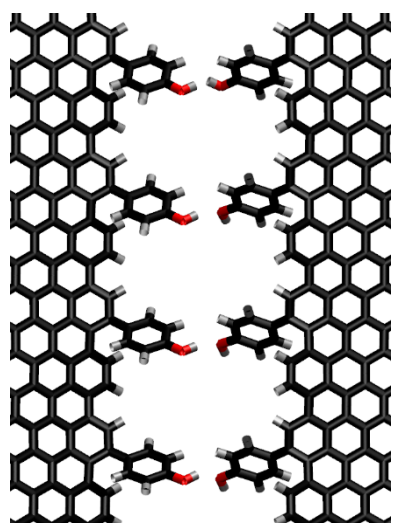

**(B)**     12.8 Å platelets distance

**Figure S14.** VMD graphical representation of phenol models. (A) Slight overlapping of the pendant phenols as observed usually up to about 9 Å of platelets distance and (B) OH alignment observed above 11 Å.

## Characterization of graphene nanopapers

**Table S3.** Density ($\rho$) and porosity ($\varphi$) of graphene nanopapers.

| Nanopaper | $\rho$ (g cm$^{-3}$) | $\varphi$ (%) |
|---|---|---|
| **GnP** | 1.30 ± 0.06 | 40.8 |
| **GnP-OH** | 1.30 ± 0.19 | 40.6 |
| **L-GnP** | 1.02 ± 0.05 | 53.6 |
| **E-GnP** | 1.14 ± 0.12 | 48.0 |

To further investigate the thermal transport and heat dissipation performances of the graphene nanopapers at the macro scale, a 15 x 4 mm$^2$ ribbons of nanopapers (30 µm thickness) made with GnP, GnP-OH or E-GnP were sandwiched horizontally between a 350 mW electrical heater and a NdFeB power magnet (3 x 3 x 3 mm$^3$) used to guarantee mechanical contact with the nanopaper, as well as to act as a thermal mass to evaluate cross plane heat exchange. The nanopaper was thus used as a heat spreader in air, while the temperature distribution as a function of time was recorded by the infrared (IR) thermal imaging camera operating in video mode (Figure S15).

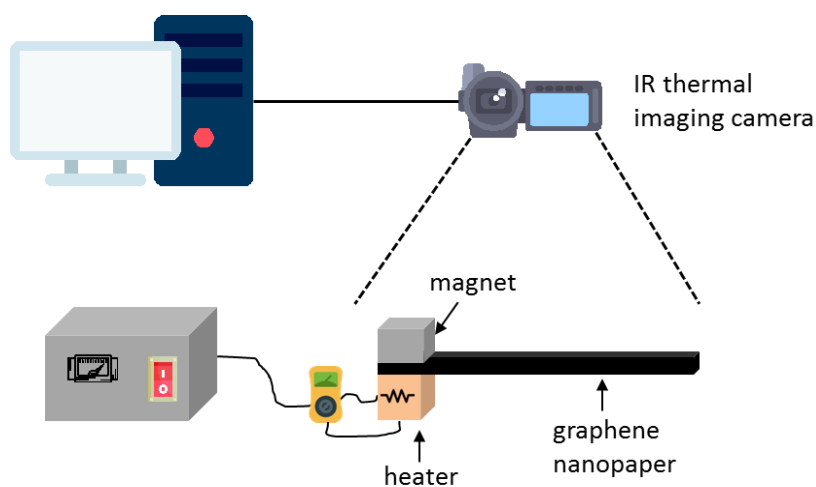

**Figure S15.** Scheme of experimental setup used to measure the temperature distribution along the nanopapers in the in-plane and through plane directions after the application of a constant power of 350 mW.

Figure S16a shows the time-sequential IR images of the samples during the heating process. The temperature on the thermal mass increases rapidly and achieve a steady-state within 4 minutes. However, significant differences were observed between the different nanopapers, as observable from the temperature profile during heating reported in Figure S16b. It can be observed that with E-GnP nanopaper a higher heating rate was obtained for the thermal mass, which reached a higher final temperature, around 73.7 °C, compared to GnP and GnP-OH, 68.4 and 69.7 °C, respectively. Hence, E-GnP confirmed a higher efficiency in the heat transfer in through-plane direction, in agreement with results obtained with the LFA. On the other hand, the thermal conductivity in the in-plane direction was investigated by analysing the temperature distribution along the length-line of the films However, temperature differences between the different nanopapers were found within the experimental errors in this setup.

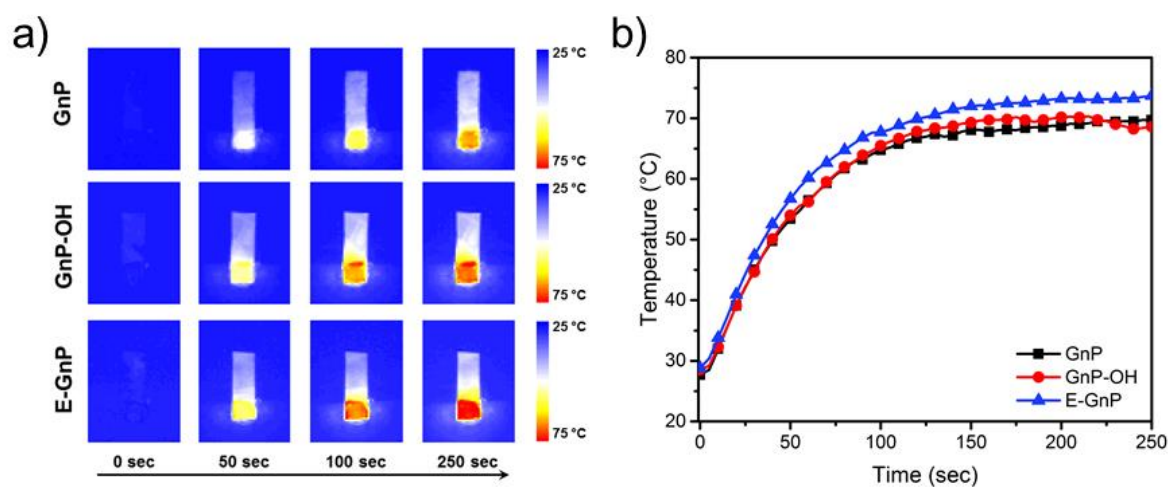

**Figure S16.** a) Time-sequential IR-camera images obtained after applying a constant current density for GnP, GnP-OH and E-GnP films and b) temperature profiles during heating.